\documentclass[conference,10pt]{IEEEtran}

\usepackage{subfigure}
\usepackage[dvips]{graphicx}
\usepackage[english]{babel}
\usepackage[latin1]{inputenc}
\usepackage{amsmath} 
\usepackage{amssymb}
\usepackage{url}

\usepackage{cite} 
\usepackage{stfloats}  
                        
\def\rien{\rule{0pt}{0pt}}
\def\Rset{\mathrm{I\!R}}

\begin{document}

\title{Application of probabilistic PCR5 Fusion Rule\\
 for Multisensor Target Tracking}
\author{%
\authorblockN{Alo\"\i s Kirchner$^a$, Fr\'ed\'eric Dambreville$^b$,\\
Francis Celeste$^c$\vspace{3pt}}
\authorblockA{D\'el\'egation G\'en\'erale pour l'Armement,\\
DGA/CEP/GIP/FAS,\\
16 bis, Avenue Prieur de la C\^ote d'Or\\
Arcueil, F 94114, France\\[3pt]
$^a$Email: alois.kirchner@polytechnique.edu\\
$^b$Form mail: http://email.fredericdambreville.com\\
$^c$Email: francis.celeste@etca.fr}
\and
\authorblockN{Jean Dezert\vspace{3pt}}
\authorblockA{ONERA,\\
Inform. Proc. and Modeling Dept.\\
29 Av. de la Division Leclerc,\\
92320 Ch\^atillon, France.\\[3pt]
Email: jean.dezert@onera.fr}
\and
\authorblockN{Florentin Smarandache\vspace{3pt}}
\authorblockA{University of New Mexico,\\
Department of Mathematics,\\
Gallup, NM 87301, USA.\\[3pt]
Email: smarand@unm.edu}}

\maketitle

\selectlanguage{english}

\begin{abstract}
This paper defines and implements a non-Bayesian fusion rule for combining densities of probabilities estimated by local (non-linear) filters for tracking a moving target by passive sensors.
This rule is the restriction to a strict probabilistic paradigm of the recent and efficient Proportional Conflict Redistribution rule no 5 (PCR5) developed in the DSmT framework for fusing basic belief assignments.
A sampling method for probabilistic PCR5 (p-PCR5) is defined.
It is shown that p-PCR5 is more robust to an erroneous modeling and allows to keep the modes of local densities and preserve as much as possible the whole information inherent to each densities to combine.
In particular, p-PCR5 is able of maintaining multiple hypotheses/modes after fusion, when the hypotheses are too distant in regards to their deviations.
This new p-PCR5 rule has been tested on a simple example of distributed non-linear filtering application to show the interest of such approach for future developments.
The non-linear distributed filter is  implemented through a basic particles filtering technique.
The results obtained in our simulations show the ability of this p-PCR5-based filter to track the target even when the models are not well consistent in regards to the initialization and real cinematic.
\end{abstract}

\noindent
{\bf Keywords: Filtering, Robust estimation, non-Bayesian fusion rule, PCR5, Particle filtering.}

%
\IEEEpeerreviewmaketitle
\section{Introduction}
Bayesian inference is a powerful principle for modeling and manipulating probabilistic information.
In many cases, Bayesian inference is considered as an optimal and legitimate rule for inferring such information.
Bayesian filters are typically regarded as optimal filters~\cite{arulampalam02tutorial,nadia, brehard}.
\\[5pt]
However, Bayesian methods need strong hypotheses, in particular about the information prior.
A degradation of the performance of Bayesian filter occurs if the filter is not correctly initialized or updated, in accordance to the models in use.
Being given a model of the system kinematic and of the measurement process, the main issue is to develop filtering methods which are sufficiently robust against the bias at the initialization as well as error in modeling. In this paper, a non-Bayesian rule for fusing the probabilistic information is proposed.
This rule, denoted p-PCR5, is the restriction to the probabilistic paradigm of the Proportional Conflict redistribution rule no.5 (PCR5) which has been proposed  in~\cite{DSmTBook2} for combining basic belief assignments.  p-PCR5 is also an extension of discrete PCR5 version to its continuous probabilistic counterpart.
\\[5pt]
PCR5 has been first established for combining evidences (i.e. discrete belief assignments) in the DSmT framework. In particular, it has been designed in order to cope with highly conflicting and uncertain information. This rule could be considered in a new probabilistic paradigm by restricting the basic belief assigment involved to only \emph{probabilistic belief assignment}\footnote{\label{footnote:1}The denomination \emph{probabilistic belief assignment} is prefered to \emph{Bayesian belief assignment}, generally used in the literature, since we consider that Probability and Bayesian inference are distinguishable notions.} and directly extended to densities of probabilities. This rule in non-Bayesian by nature. Although Bayesian techniques are widely well known and used in target tracking community (including authors works in tracking), it is interesting to see how such new approach can perform to estimate its real interest and potentiality. Surprisingly, it turns out through our works, that such approach is robust to an erroneous modeling:
in particular, it is able of maintaining multiple hypotheses, when they are too distant\footnote{\label{fnote:distance}A rigorous definition the notion of `distance' here is not so easy to establish.
This distance is essentially characterized by the distance between the means of the laws in regards to the deviations.
But we also take into account the direction of the deviation.
For example, let be given two uncertainty ellipses with high eccentricity and orthogonal orientations.
If these ellipses intersect at their extremities (the union forms a corner or a $\top$ instead of a cross), we will consider that the laws tend to be distant. 
This definition is not investigated further in the paper.} for fusion.
The resulting p-PCR5-based filter happens to be essentially non-linear, and has been implemented in our simulation using particle filtering techniques.
In particular, the p-PCR5 multisensor filter developed here is based on a quite simple and direct implementation in terms of particles drawing and resampling.
We will show the robustness of such elementary version of p-PCR5 filter, even in case of poor initialization of the filter.
\\[5pt]
Section~\ref{F2K7:section:2} introduces the PCR5 rules, and establishes some results about probabilistic PCR5.
A sampling method is deduced. 
Section~\ref{F2K7:section:3} compares the results of the Bayesian rule and of probabilistic PCR5 on a simple example.
On the basis of this comparison, some arguments about the robustness of PCR5 are given.
Section~\ref{sec:Empirical results} implements PCR5 on small tracking applications (only the filtering aspects are considered).
Distributed filtering on bearing-only sensors is considered.
Section~\ref{F2K7:sect:conclude} concludes.
%
\section{PCR5 formula for densities}
\label{F2K7:section:2}

\subsection{Definition and justification of PCR5}

The Proportional Conflict Redistribution rule no. 5 (PCR5) of combination comes from the necessity to manage precisely and efficiently the partial conflicts when combining conflicting and uncertain information expressed in terms of (quantitative) discrete belief assignments. 
It has been proved useful and powerful in several applications where it has been used~\cite{DSmTBook2}.
\\[5pt]
Let be given an universe of events $\Theta$\,.
A distribution of evidence over $\Theta$ is characterized by means of a basic belief assignment (bba) $m: \mathcal{P}(\Theta) \rightarrow \Rset^+$ such that:
$$m(\emptyset)=0\qquad \text{and} \qquad \sum_{X \subset\Theta}m(X)=1\;,$$
\noindent 
where $\mathcal{P}(\Theta)$ is the set of subset of $\Theta$.\footnote{In the general case, bba could also be defined over hyper-power sets (Dedekind's lattice)~\cite{DSmTBook2}.}
\\[5pt]
A bba typically represents the knowledge, which can be both uncertain and imprecise, that a sensor provides about its belief in the true state of the universe.
The question then arising is \emph{How to fuse the bba's related to multiple sensor responses?}
The main idea is to corroborate the information of each sensor in a conjunctive way.
\\
\emph{Example.} Let $A,B\subset \Theta$ and let's assume two sources with basic belief assignments $m_1$ and $m_2$ such that $m_1(A)=0.6,\; m_1(A\cup B)=0.4$ and $m_2(B)=0.3,\; m_2(A\cup B)=0.7$\,.
The fused bba is then characterized in a conjunctive way by:
{\footnotesize$$\begin{array}{@{}l@{}}
m_{12}(A\cap B)=m_1(A)m_2(B)=0.18\;,\\
m_{12}(A)=m_1(A)m_2(A\cup B)=0.42\;,\\
m_{12}(B)=m_1(A\cup B)m_2(B)=0.12\;,\\
m_{12}(A\cup B)=m_1(A\cup B)m_2(A\cup B)=0.28\;.
\end{array}$$}%
The conjunctive consensus works well when there is no possibility of conflict.
Now, make the hypothesis $A\cap B=\emptyset$\,.
Then, it is obtained $m_{12}(\emptyset)=0.18$, which is not an acceptable result for a conventional interpretation of $\emptyset$ as a contradiction.
Most existing rules solve this issue by redistributing the conflict $m_{12}(\emptyset)$ over the other propositions.
In PCR5, the partial conflicting mass $m_1(A)m_2(B)$ is redistributed to $A$ and $B$ only with the respective proportions $x_A=0.12$ and $x_B=0.06$\,, according to the proportionalization principle:
$$\frac{x_A}{m_1(A)}=\frac{x_B}{m_2(B)}= \frac{m_1(A)m_2(B)}{m_1(A)+m_2(B)}=\frac{0.18}{0.9}=0.2\;.$$
Basically, the idea of PCR5 is to transfer the conflicting mass only to the elements involved in the conflict and proportionally to their individual masses.
\\[5pt]
Some theoretical considerations and justifications already briefly aforementioned led to the following PCR5 combination rule.
Being given two bbas $m_1$ and $m_2$, the fused bba $m_{\text{\tiny{PCR5}}}$ according to PCR5 is defined by:
\begin{multline}
m_{\text{\tiny{PCR5}}}(X)=m_{12}(X) \\
+\sum_{\substack{Y\in \mathcal{P}(\Theta)\\ X\cap Y=\emptyset}} 
[\frac{m_1(X)^2m_2(Y)}{m_1(X)+m_2(Y)} + \frac{m_2(X)^2 m_1(Y)}{m_2(X)+m_1(Y)}]
 \label{eq:mPCR5}
 \end{multline}
\noindent
where $m_{12}(\cdot )$ corresponds to the conjunctive consensus:
$$m_{12}(X)\triangleq \sum_{\substack{X_1,X_2\in \mathcal{P}(\Theta) \\ X_1\cap X_2=X}}m_{1}(X_1)m_{2}(X_2)\;.$$
\noindent
{\small\emph{N.B.} If a denominator in~(\ref{eq:mPCR5}) is zero, the fraction is discarded.}
\subsection{Definition of probabilistic PCR5 (p-PCR5)}
\label{F2K7:PCR5:2}

In ~\cite{DSmTBook2}, Dezert and Smarandache proposed also a probabilistic version of the PCR5 rule \eqref{eq:mPCR5} by restricting the bbas $m_1$ and $m_2$ to discrete probabilities $P_1$ and $P_2$ which are called then \emph{probabilistic belief assignments/masses}$^{\ref{footnote:1}}$.
Probabilistic belief masses
are bbas, which
focal elements\footnote{Focal elements are elements of $\mathcal{P}(\Theta)$ having a strictly positive mass.} consist only in elements of the frame $\Theta$, i.e. the singletons only.
When dealing with probabilistic belief assignments $m_1\equiv P_1$ and $m_2\equiv P_2$, the PCR5 formula \eqref{eq:mPCR5} reduces to:
\begin{multline}
P_{\text{\tiny{PCR5}}}(X)=P_1(X)\sum_{Y \in \Theta} \frac{P_1(X)P_2(Y)}{P_1(X)+P_2(Y)}\\
+P_2(X)\sum_{Y \in \Theta} \frac{P_2(X)P_1(Y)}{P_2(X)+P_1(Y)}
\label{PCR5b}
 \end{multline}

\subsubsection{Extension of p-PCR5 on continuous propositions}

The previous discrete p-PCR5 formula is now extended to densities of probabilities of random variables.
Formula \eqref{PCR5b} 
is thus adapted for the fusion of continuous densities $p_1$ and $p_2$:
\begin{equation}
\label{eq:PCR5}
\begin{array}{@{}l@{}}\displaystyle
p_{12}(x)\triangleq p_{\text{\tiny{PCR5}}}(x)=p_1(x)\int_\Theta\frac{p_1(x)p_2(y)}{p_1(x)+p_2(y)} dy
\vspace{5pt}\\\displaystyle
\rien\hspace{75pt}+ p_2(x)\int_\Theta\frac{p_2(x)p_1(y)}{p_2(x)+p_1(y)} dy\;.
\end{array}
\end{equation}
%
%
\subsubsection{Properties}
In this paragraph, some properties of the (continuous) p-PCR5 are derived, which are useful for practical manipulations.
In particular, it is proved that the fused density $p_{12}$ is a true density of probability. 
\paragraph{Expectation}
The expectation of a function according to the fused probability $p_{12}$ is expressed from the initial probabilities $p_1$ and $p_2$:
\begin{equation}
\label{Ref:DMB:1}\begin{array}{@{}l@{}}\displaystyle
\int_\Theta p_{12}(y) f(y,z)\,dy=\int\!\!\int_{\Theta^2} p_1(y_1)p_2(y_2)
\vspace{5pt}\\\displaystyle
\rien\hspace{60pt}\times\frac{p_1(y_1)f(y_1,z)+p_2(y_2)f(y_2,z)}{p_1(y_1)+p_2(y_2)}dy_1dy_2
\end{array}\end{equation}
\emph{Proof.}
$$\begin{array}{@{}l@{}}
\displaystyle
\int_\Theta p_{12}(y) f(y,z)\,dy=
\int\!\!\int_{\Theta^2} \left(
\frac{p_1^2(y_1)p_2(y_2)}{p_1(y_1)+p_2(y_2)}f(y_1,z)
\right.\vspace{5pt}\\\displaystyle\left.
\rien\qquad\qquad
+
\frac{p_2^2(y_1)p_1(y_2)}{p_2(y_1)+p_1(y_2)}f(y_1,z)
\right)dy_1dy_2
\vspace{10pt}\\\displaystyle
\rien\qquad
=
\int\!\!\int_{\Theta^2}
\frac{p_1^2(y_1)p_2(y_2)}{p_1(y_1)+p_2(y_2)}f(y_1,z)
dy_1dy_2
\vspace{5pt}\\\displaystyle
\rien\qquad\qquad
+
\int\!\!\int_{\Theta^2}
\frac{p_2^2(y_1)p_1(y_2)}{p_2(y_1)+p_1(y_2)}f(y_1,z)
dy_1,dy_2
\vspace{10pt}\\\displaystyle
\rien\qquad
=
\int\!\!\int_{\Theta^2}
\frac{p_1^2(y_1)p_2(y_2)}{p_1(y_1)+p_2(y_2)}f(y_1,z)
dy_1dy_2
\vspace{5pt}\\\displaystyle
\rien\qquad\qquad
+
\int\!\!\int_{\Theta^2}
\frac{p_2^2(y_2)p_1(y_1)}{p_2(y_2)+p_1(y_1)}f(y_2,z)
dy_2dy_1\;.
\end{array}
$$
$\Box\Box\Box$\\[7pt]
\emph{Corollary.}
The density $p_{12}$ is actually probabilitic, since it is derived $\int_\Theta p_{12}(y)\,dy=1$ by taking $f=1$\,.
\paragraph{Alternative rule definition}
Let $\delta[y=z]$ be the dirac of variable $y$ over $z$.
Then:
\begin{equation}
\label{Ref:DMB:2}
\begin{array}{@{}l@{}}
\displaystyle
p_{12}(z)=\int\!\!\int_{\Theta^2} p_1(y_1)p_2(y_2) \pi(z|y_1,y_2)
\,dy_1dy_2\;,
\vspace{5pt}\\\displaystyle
\rien\quad\mbox{where}\quad
\pi(z|y_1,y_2)=\frac{p_1(y_1)\delta[y_1=z]+p_2(y_2)\delta[y_2=z]}{p_1(y_1)+p_2(y_2)}\;.
\end{array}
\end{equation}
\emph{Proof.}\\
Apply lemma 1 to the dirac distribution $f(y,z)=\delta[y=z]$\,.
\\$\Box\Box\Box$\\[7pt]
\emph{Corollary [Monte-Carlo method].}
Being able to sample $p_1$ and $p_2$, then it is possible to sample $p_{12}$ by means of the following process (let $z$ be the sample to be generated):
\begin{enumerate}
\item Generate $y_1$ according to $p_1$ and $y_2$ according to $p_2$,
\emph{together with their evaluations $p_1$ and $p_2$},
\item Generate $\theta\in[0,1]$ according to the uniform law,
\item If $\theta<\frac{p_1(y_1)}{p_1(y_1)+p_2(y_2)}$\,, then set $z=y_1$ else set $z=y_2$\,.
\end{enumerate}
It is seen subsequently that p-PCR5 is not a linear process. As a consequence, its manipulation is essentially addressed by means of Monte-Carlo method, and the previous sampling method is widely implemented in the applications.
\\[5pt]
The next section is devoted to a comparison of p-PCR5 and Bayesian rules on very simple examples.
%
\section{Bayes versus p-PCR5; whitened p-PCR5 rule}
\label{F2K7:section:3}
\subsection{Bayesian fusion rule}
In this section, we are interested in the fusion of two independent estimators by means of the Bayesian inference.
Such fusion has to take into account the prior about the state of the system.
Subsequently, this prior will be chosen to be uniform.
Although this is just a particular case of application, it will be sufficient for our purpose, \emph{i.e.} the illustration of essential differences between the Bayesian and PCR5 approaches.
\subsubsection{General case}
In Bayesian filter, the estimator is explained by means of the posterior probability $p(x|z_1,z_2)$ conditionally to the observation $z_1$ and $z_2$.
Notice that this posterior estimation should not be confounded with the true state of the system.
Now, our purpose here is to derive a rule for deriving the global estimator $p(x|z_1,z_2)$ from the partial estimators $p(x|z_1)$ and $p(x|z_2)$.
Applying Bayes' rule, one gets
$
p(x|z_1,z_2) \propto p(z_1,z_2|x)p(x)\,.
$\footnote{$p(\alpha|\beta)\propto\gamma$ means ``$p(\alpha|\beta)$ is proportional to $\gamma$ for $\beta$ fixed''.}
To go further in the derivation, it is assumed here the conditional independence between the two probabilistic sources/densities, \emph{i.e.} $p(z_1,z_2|x)=p(z_1|x)p(z_2|x)$\,.
As a consequence,
$
p(x|z_1,z_2) \propto p(z_1|x)p(z_2|x)p(x)\;,
$
and then:
\begin{equation}
\label{Bayes:fusion:eq:1}
p(x|z_1,z_2)\propto \frac{p(x|z_1)p(x|z_2)}{p(x)}\;.
\end{equation}
So, in order to compute $p(x|z_1,z_2)$, it is needed both $p(x|z_1)$, $p(
x|z_2)$ and the prior $p(x)$\,.
If one assumes uniform prior for $p(x)$,
and using notations $p_{12Bayes}=p(\cdot|z_1,z_2)$, $p_{1}=p(\cdot|z_1)$ and $p_{2}=p(\cdot|z_2)$, the Bayes' fusion formula~\eqref{Bayes:fusion:eq:1} becomes:
\begin{equation}
\label{F2K7:sectComp:eq:1}
p_{12Bayes}(x)\propto p_{1}(x)p_{2}(x)\;.
\end{equation}
\subsubsection{Gaussian subcase}

We investigate here the solution of the problem when $p_1$ and $p_2$ are Gaussian distributions.
So let's suppose for simplicity $p_1(x)$ and $p_2(x)$ mono-dimensional Gaussian distributions given by:
$$p_1(x)=\frac{1}{\sigma_1\sqrt{2\pi}} e^{-\frac{1}{2}   \frac{{(x-\bar{x}_1)}^2}{ {\sigma_1}^2 }   } \ \text{and} \ p_2(x)=\frac{1}{\sigma_2\sqrt{2\pi}} e^{-\frac{1}{2}   \frac{{(x-\bar{x}_2)}^2}{ {\sigma_2}^2 }   }$$
In absence of prior information, one assumes as usual $p(x)$ uniform.
The Bayesian rule requires to compute~(\ref{F2K7:sectComp:eq:1}).
Then, it is easily shown that $p_{12Bayes}$ is Gaussian:
$$
p_{12Bayes}(x)= \frac{1}{\sigma_{Bayes}\sqrt{2\pi}}e ^{-\frac{1}{2}\frac{{(x-\bar{x}_{bayes})}^2}{ {\sigma_{Bayes}}^2}}\;,
$$
with
$\sigma_{Bayes}^2= \frac{\sigma_1^2\sigma_2^2}{\sigma_1^2+\sigma_2^2}$
and
$\bar{x}_{Bayes}= \sigma^2_{Bayes}\left(\frac{\bar{x}_1}{\sigma_1^2} + \frac{\bar{x}_2}{\sigma_2^2}\right)\,.$
When $\sigma_1=\sigma_2=\sigma$, it is implied then:
$$\sigma^2_{Bayes}(x)= \frac{\sigma^2}{2}
\mbox{ and }
\bar{x}_{Bayes}=\frac{\bar{x}_1+\bar{x}_2}2\;.
$$
The theoretical plots and those obtained with Monte Carlo simulation are given in figures~\ref{fig1},
\ref{fig2} and~\ref{fig2part}.
These figures make the comparison with the p-PCR5 fused densities.
This comparison will be discussed subsequently.
While the Bayesian estimator is optimal (it minimizes the variance of the error estimation), it appears also that it replaces the original modes in $p_1$ and $p_2$ by a unique mode in $p_{12Bayes}$\,.
When the original modes are distant$^{\ref{fnote:distance}}$ like in figure~\ref{fig2} (for example, owing to a bad initialization of the filters), it may be interesting to keep the original modes in the fused density until it is possible to decide.
This is what p-PCR5 does.
\subsection{Fusion based on PCR5 for Gaussian distributions}
\begin{figure}
\begin{center}
\includegraphics[width=8cm]{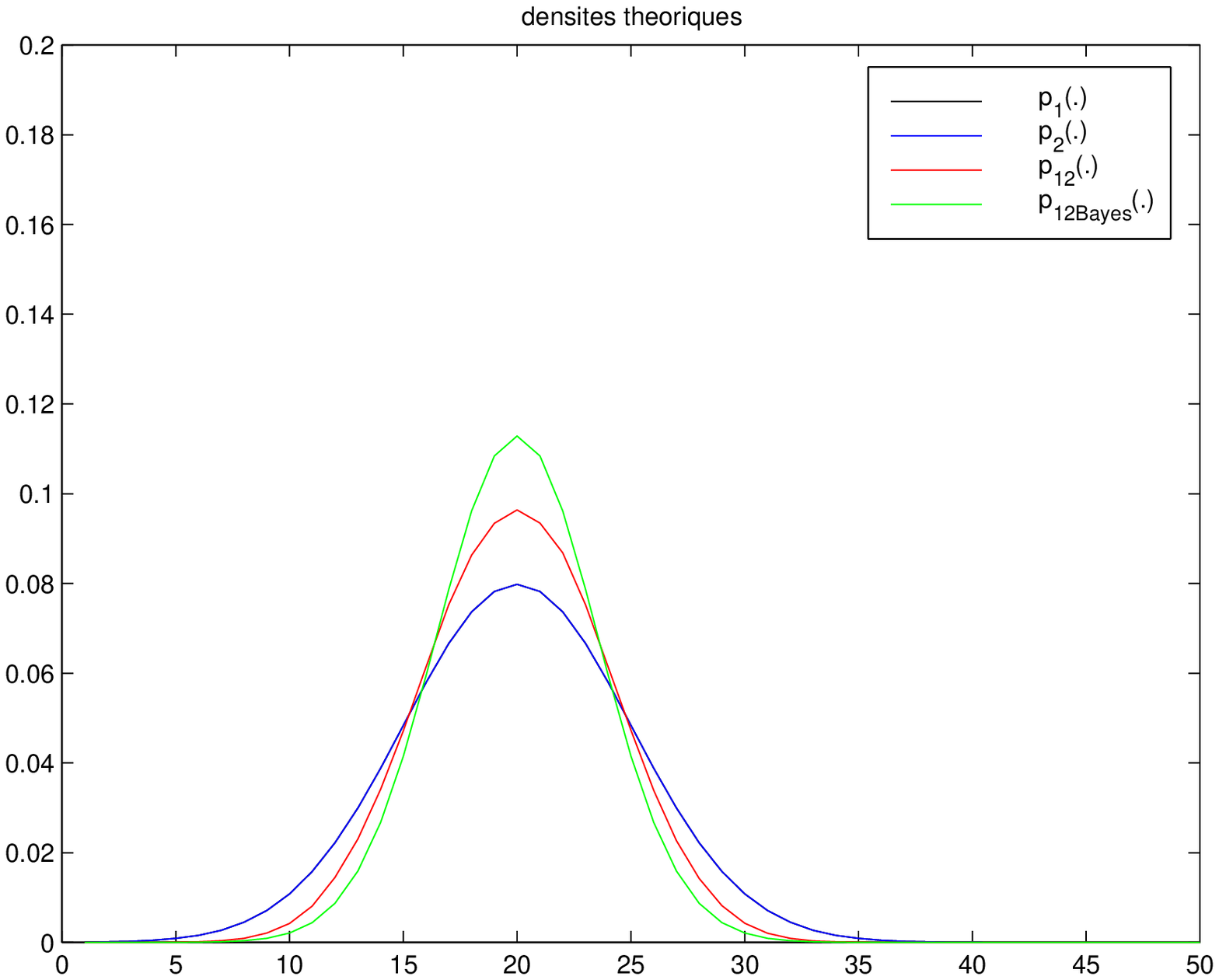}
\caption{p-PCR5 fusion versus Bayesian fusion (theoretical)}
\label{fig1}
\end{center}
\end{figure}

\begin{figure}
\begin{center}
\includegraphics[width=8cm]{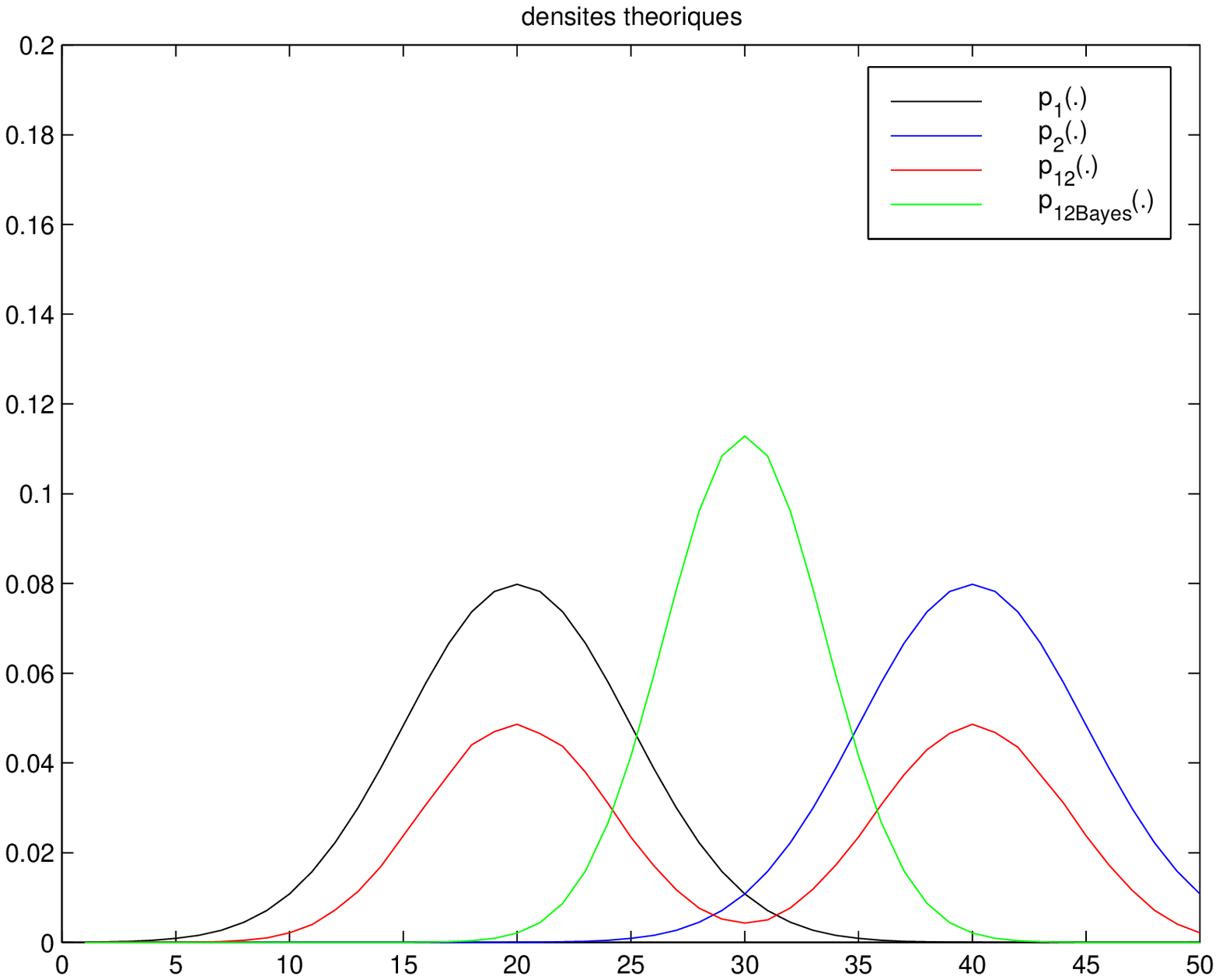}
\caption{p-PCR5 fusion versus Bayesian fusion (theoretical)}
\label{fig2}
\end{center}
\end{figure}

\begin{figure}
\begin{center}
\includegraphics[width=8cm]{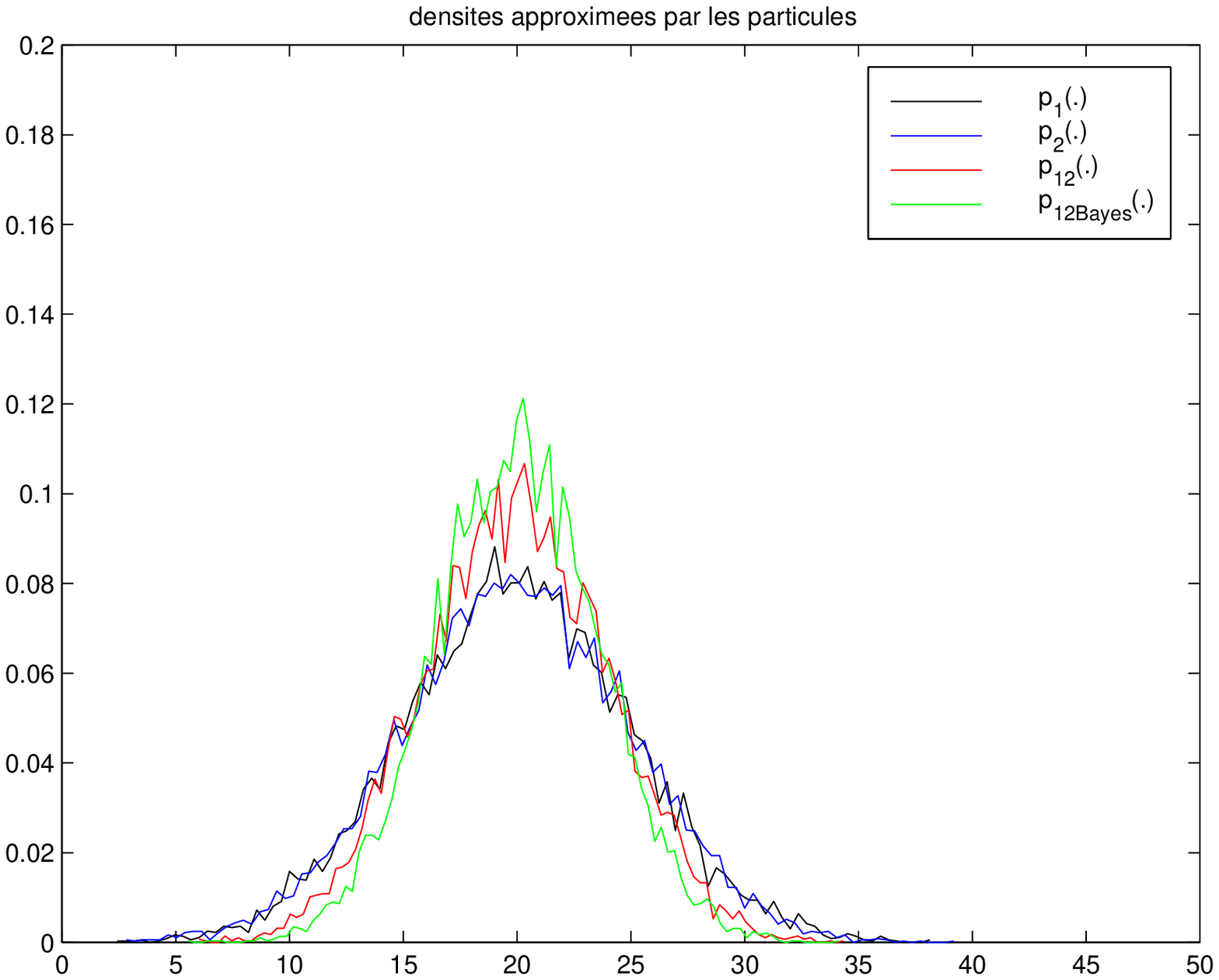}
%
%
%
\includegraphics[width=8cm]{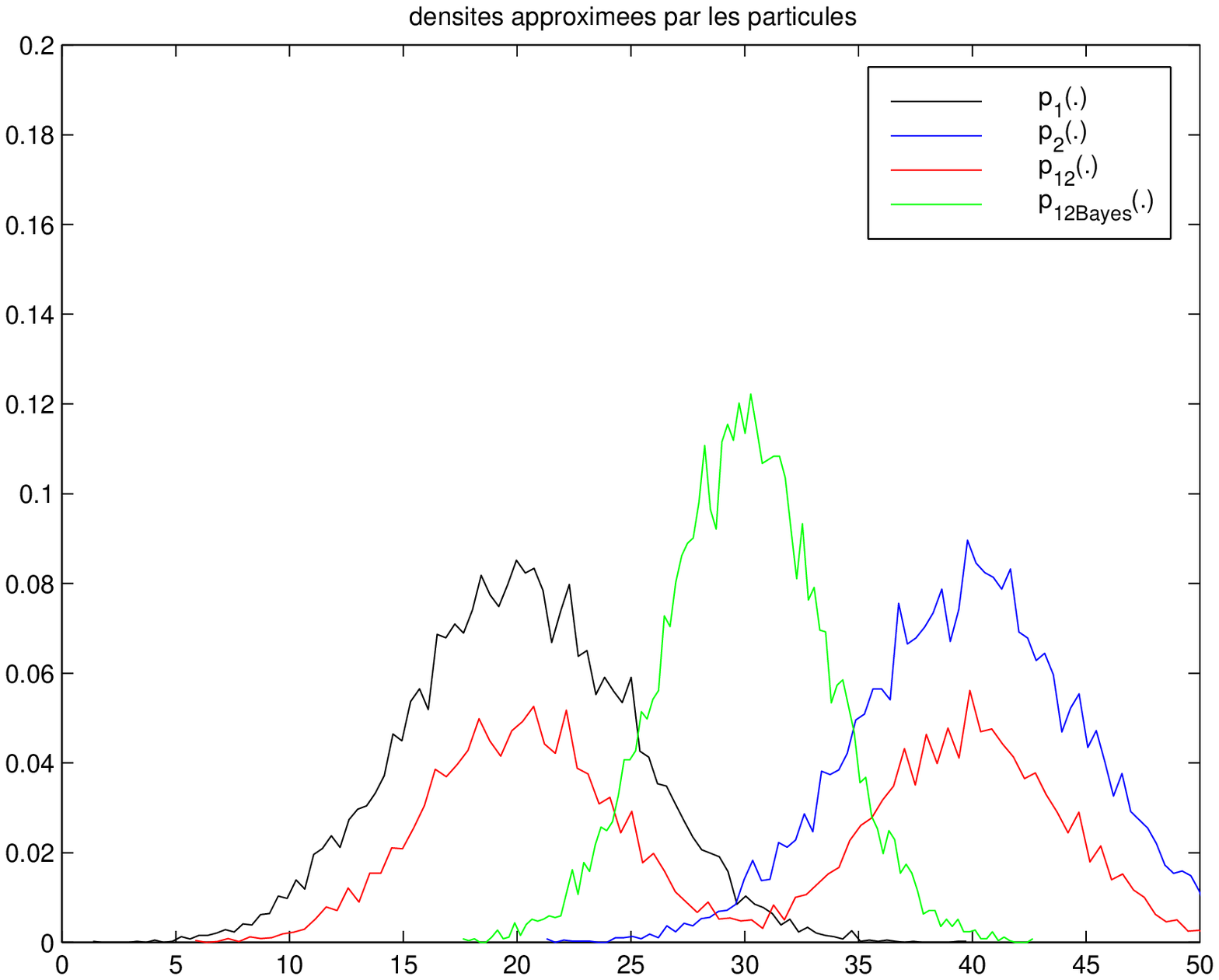}
\caption{p-PCR5 fusion versus Bayesian fusion (based on 10000 samples)}
\label{fig2part}
\end{center}
\end{figure}

The same Gaussian distribution, $p_1$ and $p_2$, are considered, but are now fused by p-PCR5 rule \eqref{eq:PCR5}, thus resulting in density $p_{12}$\,.
The fused densities are both computed, figures~\ref{fig1} and~\ref{fig2}, and sampled, figure~\ref{fig2part}.
Direct computations are expensive, and are obtained in two steps:
\begin{itemize}
\item Compute
\mbox{$ I_s(x)=\int \frac{p_s(x)p_{\bar s}(y)}{p_s(x)+p_{\bar s}(y)} dy$}, where $s\in\{1,2\}$ and $\bar s\in\{1,2\}\setminus\{s\}$\,,
\item Then compute $p_{12}(x)=p_1(x) I_1(x) + p_2(x) I_2(x)$\,.
\end{itemize}
It appears clearly that computed and sampled densities match well, thus confirming the rigthness of our sampling method.
Now, contrariwise to the Bayesian rule, it is noticed two different behaviors (which are foreseeable mathematically):
\begin{itemize}
\item When the densities $p_1$ and $p_2$ are close$^{\ref{fnote:distance}}$, $p_{12}$ act as an amplifier of the information by reducing the variance.
However, this phenomena is weaker than for $P_{12Bayes}$.
p-PCR5 is thus able to amplify the fused information, but is less powerful than the Bayesian rule in this task.
\item When the densities $p_1$ and $p_2$ are distant$^{\ref{fnote:distance}}$, $p_{12}$ keeps both modes present in each density and preserves the richness of   information by not merging both densities into only one (unimodal) Gaussian density. This is a very interesting and new property from a theoretical point of view, which presents advantages for  practical applications as shown in the following simple tracking example.
\end{itemize}
In regards to these differences, it is thus foreseeable that the p-PCR5 should be more robust to potential errors.
\subsection{Whitened p-PCR5 rule}
It has been seen that the p-PCR5 fusion of the same densities $p_1=p_2$ will result in an amplified density $p_{12}$.
Of course, this is not practicable when the densities $p_1$ and $p_2$ are related to correlated variables.
Consider for example that the state $y$ are measured by $z^1$ and $z^2$.
The (distributed) posterior probabilities are $p_s(y)=p(y|z^s)\propto p(y)p(z^s|y)$ for $s=1,2$\,.
It happens that the variables estimated by $p_1$ and $p_2$ are correlated, so that p-PCR5 should not be applied directly.
In particular, the fusion of $p_1$ and $p_2$ by means of p-PCR5 results in a density $p_{12}$ stronger than the prior $p$ over $y$, even when there is no informative measure, \emph{i.e.} $p(z^s|y)=p(z^s)$\,!
In order to handle this difficulty, we propose a \emph{whitened} p-PCR5 rule, producing a fused density $p_{\text{\tiny{whitePCR5}}}$ from the updated information only:
\begin{equation}
\label{Ref:DMB:white:2}
\begin{array}{@{}l@{}}
\displaystyle
p_{\text{\tiny{whitePCR5}}}(y)=\int\!\!\int_{\Theta^2} p_1(y_1)p_2(y_2) \pi(y|y_1,y_2)
\,dy_1dy_2\;,
\vspace{5pt}\\\displaystyle
\mbox{where}\ 
\pi(y|y_1,y_2)=\frac{
\frac{p(y_1|z^1)}{p(y_1)}\delta[y_1=y]+\frac{p(y_2|z^2)}{p(y_2)}\delta[y_2=y]}{\frac{p(y_1|z^1)}{p(y_1)}+\frac{p(y_2|z^2)}{p(y_2)}}\;.
\end{array}
\end{equation}
In~\eqref{Ref:DMB:white:2}, the proportion $\frac{p(y|z^s)}{p(y)}$ should be considered as the information intrinsically obtained from sensor $s$.
It happens that the whitened p-PCR5 does not change the prior when there is no informative measure, \emph{i.e.} $p_{\text{\tiny{whitePCR5}}}(y)=p(y)$ when $p(z^s|y)=p(z^s)$ for $s=1,2$\,.  

\section{A distributed sequential filtering application}
\label{sec:Empirical results}
\subsection{Theoretical setting}
A target is moving according to a known Markov prior law.
Let $y_t$ be the state of the target at time $t$.
It is assumed:
$$
p(y_{1:t+1})=p(y_{t+1}|y_t)p(y_{1:t})\;.
$$
In order to estimate the state of the target, $S=2$ sensors are providing some measurements.
Denote $z^s_t$ be the measurement of the state $y_t$ by sensor $s$.
The measure is characterized by the law $p(z^s_t|y_t)$, which is known.
It is assumed that the measure are made independently, conditionally to a given state:
$$
p(z^{1:S}_t|y_t)=\prod_{s=1}^Sp(z^s_t|y_t)\;.
$$
Our purpose is to derive or approximate the optimal estimator, $p(y_{t+1}|z^{1:S}_{1:t+1})$\,, from the distributed retroacted estimators, $p(y_{t+1}|z^{1:S}_{1:t},z^{s}_{t+1})$, related to sensors $s$.
There is a Bayesian approach to this problem, and we propose some comparison with a p-PCR5 approach and a whitened p-PCR5 approach.
\subsubsection{Distributed Bayesian filter}
It is derived from: 
{\small\begin{eqnarray}
\label{distri:Bayes:filter:eq:0:1}
p(y_{t:t+1}|z^{1:S}_{1:t})= p(y_{t+1}|y_t)p(y_{t}|z^{1:S}_{1:t})
\;,\\[3pt]
\label{distri:Bayes:filter:eq:1}
p(y_{t:t+1}|z^{1:S}_{1:t},z^{s}_{t+1})\propto p(z^{s}_{t+1}|y_{t+1})p(y_{t:t+1}|z^{1:S}_{1:t})
\;,\\[3pt]
\label{distri:Bayes:filter:eq:2}
p(y_{t:t+1}|z^{1:S}_{1:t+1})\propto\biggl(\prod_{s=1}^S\frac{p(y_{t+1}|z^{1:S}_{1:t},z^{s}_{t+1})}{p(y_{t+1}|z^{1:S}_{1:t})}\biggr)
p(y_{t:t+1}|z^{1:S}_{1:t})
\;.
\end{eqnarray}}%
This approach is unstable, when some components of the target state are non-observable; for example, adaptations of the method are necessary~\cite{brehard} for bearing only sensors.
However, the method will be applied as it is here to bearing only sensors, in order to compare to the robustness of the PCR5 approach.  

\subsubsection{p-PCR5 filter}
It is derived from~\eqref{distri:Bayes:filter:eq:0:1},~\eqref{distri:Bayes:filter:eq:1} and: 
{\small\begin{equation}
\label{distri:p-PCR5:filter:eq:3}
\begin{array}{@{}l@{}}
\displaystyle
p(y_{t+1}|z^{1:S}_{1:t+1})=\!
\int_{y^{1:S}_{t+1}}\!\!\biggl(\prod_{s=1}^Sp(y^s_{t+1}|z^{1:S}_{1:t},z^{s}_{t+1})\biggr)
\pi(y_{t+1}|y^{1:S}_{t+1})
dy^{1:S}_{t+1}
\vspace{5pt}\\\displaystyle
\mbox{where}\ 
\pi(y_{t+1}|y^{1:S}_{t+1})=\frac{\sum_{s=1}^Sp(y^s_{t+1}|z^{1:S}_{1:t},z^{s}_{t+1})\delta[y_{t+1}=y^s_{t+1}]}{\sum_{s=1}^Sp(y^s_{t+1}|z^{1:S}_{1:t},z^{s}_{t+1})}
\;,
\end{array}
\end{equation}}%
and $p(y^s_{t+1}|z^{1:S}_{1:t},z^{s}_{t+1})$ is an instance of $p(y_{t+1}|z^{1:S}_{1:t},z^{s}_{t+1})$, obtained by just replacing $y_{t+1}$ by $y^s_{t+1}$\,.
\\
It is noticed that this filter is necessary suboptimal, since it makes use of the p-PCR5 rule on correlated variables.
More precisely, the outputs $y^s_{t+1}$ of the local filters are in fact related to the same prior estimation at time $t+1$.
The fusion without correction by p-PCR5 implies a redundancy of the prior estimation.
The whitened p-PCR5 filter will resolve this difficulty.
By the way, it is seen that the p-PCR5 filter still works experimentally on the considered examples.
\subsubsection{Whitened p-PCR5 filter}
It is derived from~\eqref{distri:Bayes:filter:eq:0:1},~\eqref{distri:Bayes:filter:eq:1} and: 
{\small\begin{equation}
\label{distri:p-PCR5:filter:eq:4}
\begin{array}{@{}l@{}}
\displaystyle
p(y_{t+1}|z^{1:S}_{1:t+1})=\!
\int_{y^{1:S}_{t+1}}\!\!\biggl(\prod_{s=1}^Sp(y^s_{t+1}|z^{1:S}_{1:t},z^{s}_{t+1})\biggr)
\pi(y_{t+1}|y^{1:S}_{t+1})
dy^{1:S}_{t+1}
\vspace{5pt}\\\displaystyle
\mbox{where}\ 
\pi(y_{t+1}|y^{1:S}_{t+1})=\frac{\sum_{s=1}^S\frac{p(y^s_{t+1}|z^{1:S}_{1:t},z^{s}_{t+1})}{p(y^s_{t+1}|z^{1:S}_{1:t})}\delta[y_{t+1}=y^s_{t+1}]}{\sum_{s=1}^S\frac{p(y^s_{t+1}|z^{1:S}_{1:t},z^{s}_{t+1})}{p(y^s_{t+1}|z^{1:S}_{1:t})}}
\;.
\end{array}
\end{equation}}
Again, $y^s_{t+1}$ is just an instance of $y_{t+1}$ for sensor $s$\;.
\rien\\[5pt]
These filters have been implemented by means of particles.
The sampling of p-PCR5 has been explained yet, but it is not the purpose of this paper to explain all the theory of particle filtering;
a consultation of the literature, \emph{e.g.} \cite{Ristic2004}, is expected.
\subsection{Scenario and tests}
\subsubsection{Scenario and simulation results for passive multi-sensor target tracking}

In order to test the p-PCR5 fusion rule, we simulate the following scenario:
in a 2-dimensional space, two independent passive sensors are located in (0,100) and (100,0) in Cartesian coordinates.
These sensors provide a noisy azimuth measurement (\emph{0.01 rad.} normal noise) on the position of a moving target.
We associate a tracking particle filter to each sensor.
The motion model is the following :
{\small\begin{equation}
\label{motionmodel}
\begin{array}{@{}l@{}}
\dot{x}_{t+1} = \dot{x}_{t} + 0.1*N(0,1) 
\\
\dot{x}_{t+1} = \dot{x}_{t} + 0.1*N(0,1)
\\
x_{t+1} = x_{t} + dt*\dot{x}_{t} + 0.3*N(0,1)
\\
y_{t+1} = y_{t} + dt*\dot{y}_{t} +0.3*N(0,1) 
\end{array}
\end{equation}}
\small{\centering{\emph{where dt = 1 time unit and N(0,1) is the normal distribution.}}}\\
In our simulations, each local particle filter is implemented by means of 200 particles.
At every time step, we proceed to the fusion of the local posterior densities and then re-inject the fused state density into each local filter (feedback loop).
Three different paradigms are considered for the fusion: Bayesian, p-PCR5 and whitened p-PCR5 rules.
These filters try to estimate both the mobile position and speed of the target which is assumed to follow a quasi-constant velocity model.
It is noticed that we are dealing directly with both the observable and non-observable components of the target state (each sensor is concerned).
In particular, notice that this infers an additional perturbation to the basic particle Bayesian filter, even if there is a feedback.

\subsubsection{A simple example}

\begin{figure}[hbp]
\begin{center}
\includegraphics[width=0.45\textwidth]{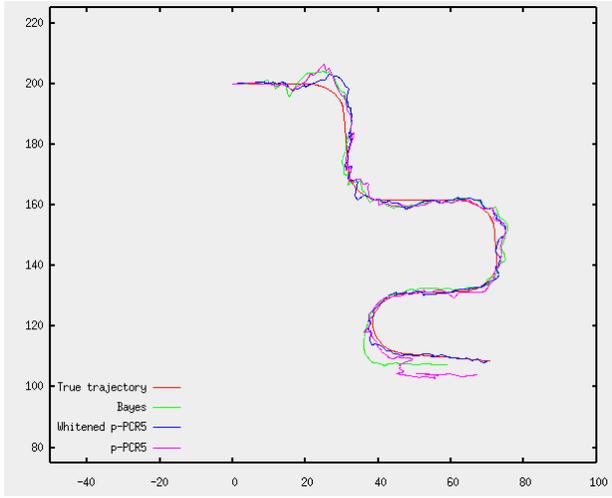}
\end{center}
\caption{Averaged trajectories using different tracking methods.} \label{traj_courbe}
\end{figure}

\begin{figure}[htbp]
\centering
{
\subfigure[Timestep 160]{\includegraphics[width=0.45\textwidth]{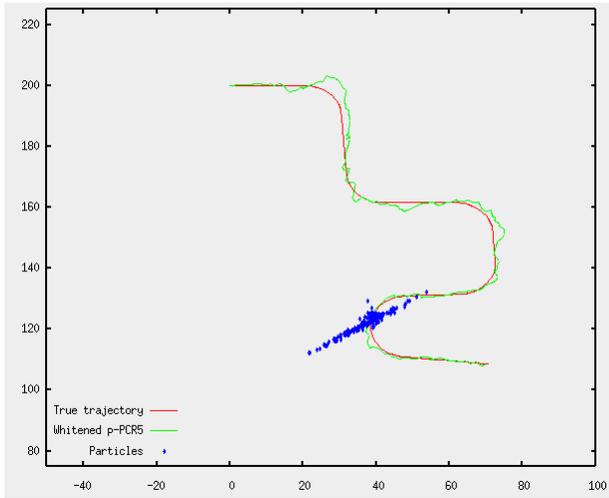}
\label{tourne}}
\\
\subfigure[Timestep 170]{\includegraphics[width=0.45\textwidth]{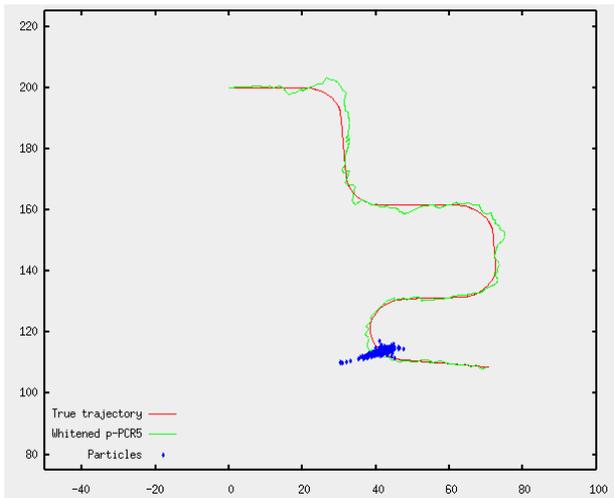}
\label{tourne2}}
}
\caption{Particle clouds for whitened p-PCR5 in the last curve.} \label{tournant}
\end{figure}

\begin{figure}[hbp]
\begin{center}
\includegraphics[width=0.45\textwidth]{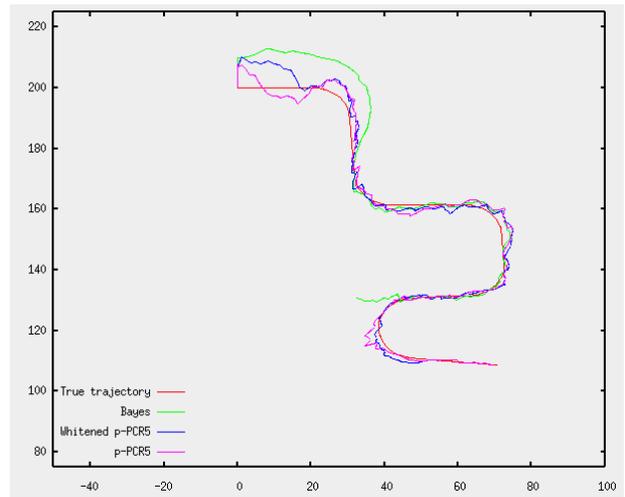}
\end{center}
\caption{Averaged trajectories using different tracking methods. Poor initialization : null speed and 10 units away starting position.} \label{traj_courbe_poor}
\end{figure}

In this first example, the filters are well initialized (we give them good starting speed and position).
The mobile follows a non-linear trajectory (figure~\ref{traj_courbe}), in order to show the capability of this distributed filter to converge.
On this example, the Bayesian filter manages to track the target with some difficulties during the last curve in figure~\ref{traj_courbe}.
On the same example, p-PCR5 and whitened p-PCR5 rules have been tested with success.
While both filters have to reestimate the speed direction at each turn, it appears that this reestimation is more difficult for p-PCR5.
This difference is also particularly apparent during the last curve.
Figure~\ref{tournant} displays the particle cloud of the whitened PCR-5 filter during and after the last curve.
The variance rises during the curve, resulting in the cross-like cloud of sub-figure~\ref{tourne}, which is characteristic to the p-PCR5 fusion: \emph{the branches correspond to the direction the sensors are looking at.}
Then, the p-PCR5, by amplifying the zone where the filters are according to see the object, allows the process to converge again toward the object real position in an expansion-contraction pattern (figure~\ref{tourne2}).
In more difficult cases, with poor initialization for instance (see figure~\ref{traj_courbe_poor}), both p-PCR5 and whitened p-PCR5 manage to follow the target, while the Bayesian filter diverges in about $33$ percent of the cases.

Next sections investigate more thoroughly the properties of the whitened p-PCR5 filtering.

\subsubsection{Whitened p-PCR5 robustness against poor initialization}

In order to test the capability of (whitened) p-PCR5 to recover from erroneous measurements of the local estimations, we considered two scenarios in which the filters are differently and badly initialized.
In these scenarios, the real trajectory of the object is the same: it starts at $(200,0)$ and moves toward $(200,150)$ at a constant speed $(0,1)$.
\begin{table}[htbp]
\renewcommand{\arraystretch}{1.3}
\caption{Initialization data}
\label{init_data}
\begin{center}
\begin{tabular}{|c|c|c|c|c|c|}
\hline
 & &x&y&x speed&y speed\\
\hline
First&Filter 1&190&10&0&0\\
\cline{2-6}
example&Filter 2&210&10&0&0\\
\hline
Second&Filter 1&190&10&0.1&-1\\
\cline{2-6}
example&Filter 2&210&10&0.5&1.5\\
\hline
\end{tabular}
\end{center}
\end{table}
\begin{figure}[htbp]
\centering
{
\subfigure[Timestep 1]{\includegraphics[width=0.33\textwidth]{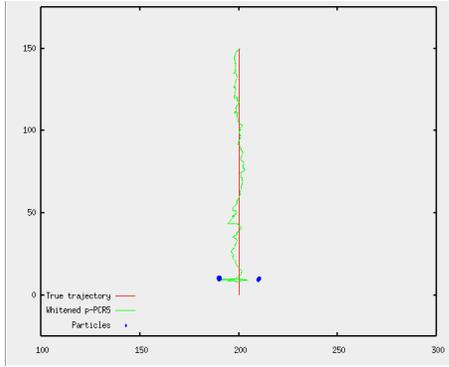}
\label{bad_init1}}
\\
\subfigure[Timestep 10]{\includegraphics[width=0.33\textwidth]{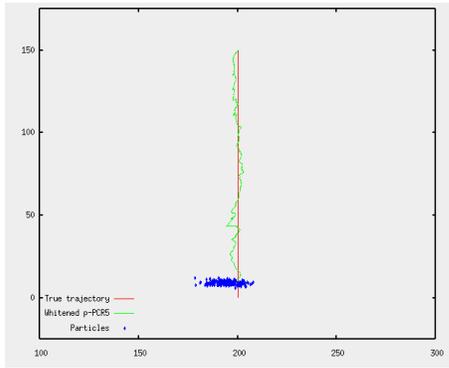}
\label{bad_init10}}
\\
\subfigure[Timestep 20]{\includegraphics[width=0.33\textwidth]{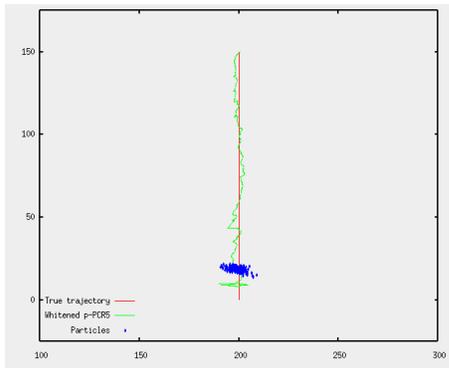}
\label{bad_init20}}
\\
\subfigure[Timestep 60]{\includegraphics[width=0.33\textwidth]{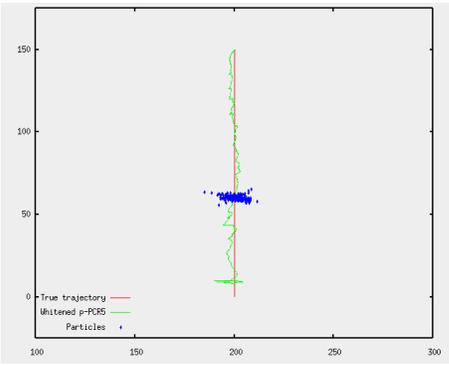}
\label{bad_init60}}
}
\caption{The real mobile starts at $(200,0)$ and moves upward at constant speed $(0,1)$; poor filters initialization.} \label{bad_init}
\end{figure}

In the first scenario (figure~\ref{bad_init}), the first filter, which sensor is placed at $(0,100)$, is initialized at position $(190,10)$ and at speed $(0,0)$.
The second filter, which sensor is at $(100,0)$, is initialized at position $(210,10)$ and at the same speed (figure~\ref{bad_init1}).
As the estimated positions are far from the real one (in regards to the noise models) and both sensors are looking at the object from a remote position, the particle cloud quickly spread horizontally (figure~\ref{bad_init10}).
Then the (whitened) PCR5 begins to find zones where both filters estimate a non-negligible probability of presence and amplifies them until convergence (figure~\ref{bad_init20}). Though the particle cloud still seems to be fairly spread (because of sensors remote position), the global estimate is very close from the real position and speed, and will remain so until the last time step (figure~\ref{bad_init60}).
\begin{figure}[htbp]
\centering
{
\subfigure[Timestep 1]{\includegraphics[width=0.33\textwidth]{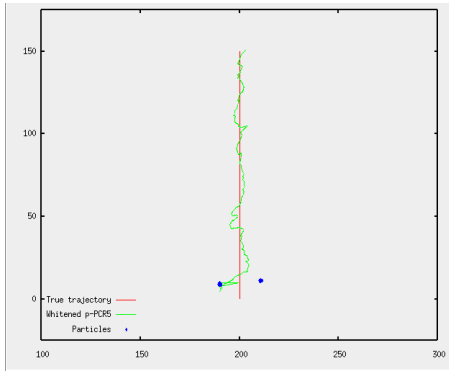}
\label{worse_init1}}
\\
\subfigure[Timestep 10]{\includegraphics[width=0.33\textwidth]{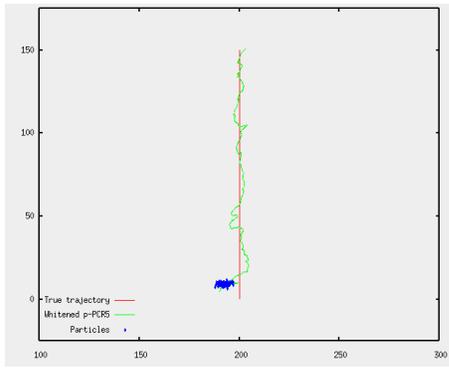}
\label{worse_init10}}
\\
\subfigure[Timestep 20]{\includegraphics[width=0.33\textwidth]{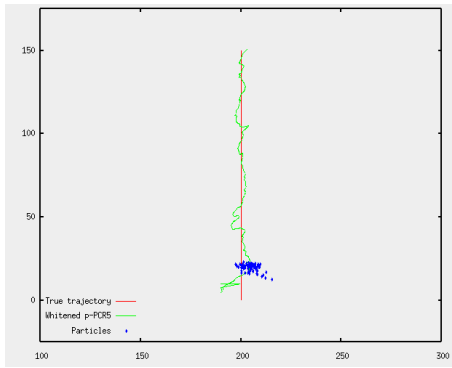}
\label{worse_init20}}
\\
\subfigure[Timestep 60]{\includegraphics[width=0.33\textwidth]{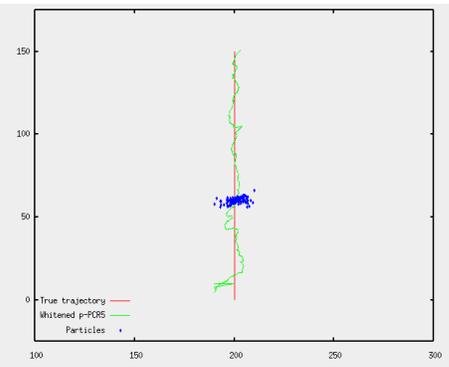}
\label{worse_init60}}
}
\caption{The real mobile starts at $(200,0)$ and moves upward at constant speed $(0,1)$; \emph{bad} filters initialization.} \label{worse_init}
\end{figure}

Our second example (figure~\ref{worse_init}) is a limit case: the initialization is quite worse (see table~\ref{init_data}), since our motion model assumes nearly constant speed and therefore makes it hard to recover from such erroneous and contradictory speed initialization.
An interesting point is that, for a tight prediction noise, p-PCR5 sometimes does not converge on this example, while whitened p-PCR5 usually does.
Artificially raising the prediction noise solves this problem for `standard' p-PCR5, showing its trend to over-concentrate the particle cloud.

\subsubsection{Whitened p-PCR5 versus mean} 

\begin{figure}[htbp]
\centering{
\subfigure[Timestep 1]{\includegraphics[width=0.3\textwidth]{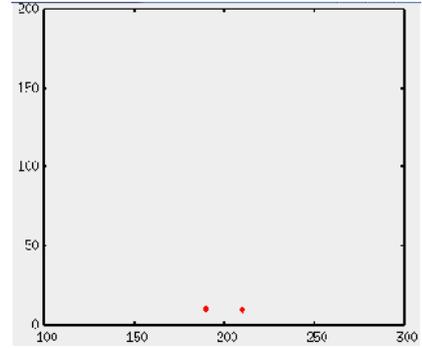}
\label{mean1}}
\\
\subfigure[Timestep 30]{\includegraphics[width=0.3\textwidth]{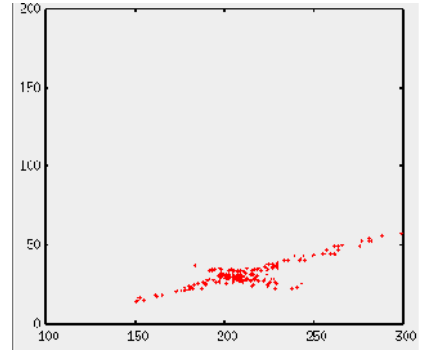}
\label{mean30}}
\\
\subfigure[Timestep 45]{\includegraphics[width=0.3\textwidth]{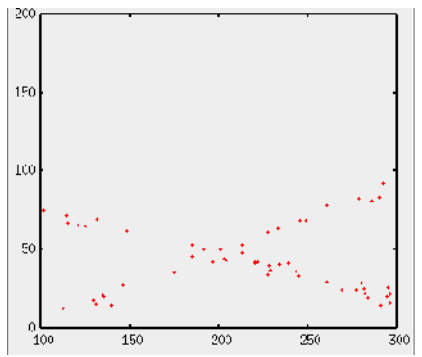}
\label{mean45}}
}
\caption{Using mean instead of p-PCR5. Red dots are the positions of the particles after fusion. The real mobile starts in (200,0) at time step 0 and moves at a constant speed of (0,1).} \label{mean}
\end{figure}

As seen before, the PCR5-fusion of two probabilistic densities amplifies the areas where both densities have a non-negligible value.
Otherwise, it usually works like just averaging the two densities.
In order to measure the impact of the amplification, we reprocessed the first example of previous subsection while using the mean, $p_{12}=\frac{p_1+p_2}{2}$, instead of p-PCR5.
The result (figure~\ref{mean}) is self explanatory: the same expansion as with PCR5 occurs (figure~\ref{bad_init}), but contraction never appears.
\subsubsection{Conclusions}
The results presented here have clearly shown that p-PCR5, and especially whitened p-PCR5, filters are more robust than the basic Bayesian filter.
However, it is clear that Bayesian filters are the best, when the priors are correctly defined and the variables are locally observable (notice that there are adaptations of the basic Bayesian filter to non-observable variables~\cite{brehard}).
The real interest of p-PCR5 is that it does not need prior knowledges about the antedating local particle filters: just apply the method and obtain consistent results!
\section{Conclusions}
\label{F2K7:sect:conclude}
This paper has investigated a new fusion rule, p-PCR5, for fusing probabilistic densities.
This rule is derived from the PCR5 rule for fusing evidences.
It has a simple interpretation from a sampling point of view.
p-PRC5 has been compared to the Bayesian rule on a simple fusion example.
Then, it has been shown that p-PCR5 was able to take into account multiple hypotheses in the fusion process, by generating multiple modes.
Thus, more robustness of p-PCR5 were foreseeable in comparison to Bayes' rule.
This robustness has been tested successfully on examples of distributed target tracking.
It is expected that this new rule will have many applications, in particular in case of filtering with incomplete models.

\end{document}